\pdfoutput=1
\documentclass{article}
\usepackage[dvipsnames]{xcolor}
\usepackage[utf8]{inputenc} 
\usepackage[T1]{fontenc}
\usepackage{hyperref}
\usepackage{mathrsfs}
\usepackage{amssymb}
 \hypersetup{
    pdftoolbar=true,        
    pdfmenubar=true,      
    pdffitwindow=false,    
    pdfstartview={FitH},
    pdfauthor={Sabina Yeasmin},    
    colorlinks=true,      
    linkcolor=blue,        
    citecolor=blue,      
}
\usepackage{url}      
\usepackage{setspace}
\usepackage[pdftex]{graphicx}
\usepackage{dcolumn}
\usepackage[margin=1in]{geometry}
\usepackage{amsmath}
\usepackage{graphicx}
\usepackage{float}
\usepackage{tabularx}
\usepackage[mathscr]{eucal}

\usepackage{caption}
\usepackage{subcaption}
\usepackage{multirow}
\usepackage{booktabs}
\usepackage[sort, numbers]{natbib}

\begin{document}
\title{Embedding Warm Natural Inflation in $f(\phi)T$ gravity}
\author{ Sabina Yeasmin \footnote{Email:sabyes22@gmail.com} , Atri Deshamukhya  \footnote{Email: atri.deshamukhya@aus.ac.in}\\
\textit{\small Department of Physics, Assam University, Silchar, Assam 788011, India}} 
\date{}

\maketitle

  \begin{center}
  \begin{minipage}{0.85\textwidth}
  
{ {\small   We study warm inflation in the framework of $f(\phi)T$ gravity, where $\phi$ is the inflaton and $T$ is the trace of the energy-momentum tensor. The inflaton field is assumed to roll on the natural potential and the result is analyzed in light of Planck 2018 and BICEP/Keck 2021 data. We start our work by obtaining the field equations under slow-roll approximations. We then evaluate the scalar and tensor power spectra and their corresponding spectral index and tensor-to-scalar ratio with a temperature-dependent form of the dissipation coefficient during the inflationary era. We find that the warm inflation model in $f(\phi)T$ gravity is compatible with observational bands.}  }
    
    \end{minipage}        
    \end{center} 
    
  \vspace{.2cm}
 \section{Introduction}
According to the Standard Model of Cosmology and observational data, the universe has undergone two phases of accelerated expansion throughout its evolution, the inflationary phase at its early stage and a second phase of accelerated expansion in recent times. This second phase of accelerated cosmic expansion is believed to be a consequence of the presence of dark energy exerting negative pressure. General Relativity is not ideal for explaining the existence of the dark sector i.e. dark matter and dark energy of the universe \cite{r1, r2, r3}. To overcome such shortcomings researchers started studying the different higher-order extensions to General Relativity and exploring alternative theories of gravity \cite{ r4, r5, r6, r7, r8}. Modified theories of gravity not only address the late-time acceleration but these theories are also applied to explain the early Universe, e.g., the inflationary era. 

In modified gravity theories, the Einstein-Hilbert action is modified either by extending the geometric or the matter part or both. Unlike Einstein's theory, whose field equations contain only up to second-order derivatives, the modified theories with higher derivative Ricci/Riemann tensor gravity models include higher derivatives \cite{r9}. A significant difference is therefore expected between general relativity and modified gravity predictions.

The $f(R)$ gravity model is the first proposed modification of the Einstein gravity where the curvature scalar in the Einstein-Hilbert action has been replaced by a general function of Ricci scalar $R$. This theory describes both inflation and late-time expansion of the universe. The first viable $f(R)$ model to study inflation was proposed by Nojiri and his collaborators \cite{r10, r11, r12, r13, r14}. T. Harko et al. \cite{r15} introduced another modified gravity theory called $f(R, T)$ theory where the gravitational field is assumed to be coupled to the trace T of the energy-momentum tensor of the matter in the action. Later several inflationary models have been studied within the context of the  $f(R, T)$ gravity theory \cite{r16, r17, r18, r19, r20, r21}. There are many other technically allowed modified gravity theories that have been investigated in literature in the context of inflationary cosmology like $f(\mathscr{T})$ \cite{r22}, $f(R, G)$ \cite{ r23}, Gauss-Bonnet (GB) theory ($f(G)$ gravity) \cite{r24}, Brans-Dicke theory \cite{r25}, etc. where $\mathscr{T}$ denotes the Torsion scalar and $G$ is Gauss-Bonnet scalar.

Another recent modification is the $f(\phi)T$ theory of gravity, proposed by Zhang et al. \cite{r26}, which is an extension of the simplest $f(R,T)$ gravity theory. This modified theory of gravity has been investigated by adding a coupling term between the scalar field $\phi$ and the trace of the energy-momentum tensor into the action in general relativity. They studied cold inflationary dynamics in this framework by considering three well-motivated potentials: chaotic, natural, and Starobinsky, and found that chaotic and natural inflationary models are in better agreement with the observational constraints. However, a larger value of the tensor-to-scalar ratio was obtained for the Starobinsky model. Although natural inflation in $f(\phi)T$ gravity is consistent with the observational data, a major problem with that is a larger axion decay constant is required to agree with the observational constraints. So, it is difficult to embed this model in fundamental theories like string theory \cite{r27}. In this study, we show that it is possible to lower the value of the axion decay constant below the Planck scale if the natural inflationary model in $f(\phi)T$ theory of gravity is studied in the context of the warm inflationary scenario.

In 1995, Berera \cite{r28} introduced the idea of warm inflation as an alternative approach to resolving the graceful exit problem of the standard inflation dynamics.
 The basic concept of warm inflation is that during inflationary period, the inflaton field does interact with other existing fields \cite{r29, r30, r31, r32, r33, r34} which leads to vacuum energy decay to the other lighter fields, thereby creating particles simultaneously with the process of inflation. Thus in warm inflation, radiation is produced simultaneously with the expansion of the universe and there can be a smooth transition from the inflationary phase to the radiation-dominated phase without the requirement of invoking any separate reheating phase. In this scenario, dissipative effects play an important role during inflation.


  In warm inflation scenario, energy dissipates from the inflaton field into radiation during the inflationary phase and this mechanism is parameterized by a term $\Gamma$ called dissipation coefficient \cite{r35, r31, r36}. Because of this dissipative coupling between inflaton and radiation field inflation can last long even if the potential is not very flat. In this work, we use cubic temperature-dependent dissipation coefficient which affects the prediction of cosmological observables like it suppresses the value of the tensor-to-scalar ratio. The latest Cosmic Microwave Background (CMB) observations of BICEP/Keck \cite{r37} combined with the Planck 2018 data \cite{r38} sets an  bound on the tensor-to-scalar ratio of $r<0.036$ at the $95\%$ confidence level. We show that the warm natural inflation in $f(\phi)T$ gravity model is consistent with this observational bound on $r$.

The paper is organized as follows: In section 2, we present a brief overview of $f(\phi)T$ gravity in the FLRW background. In section 3, we investigate warm inflation in $f(\phi)T$ gravity and study the cosmological perturbations originating from $f(\phi)T$ gravity during warm inflation and also formulate the corresponding power spectrum, the tilt of spectral index, and tensor-to-scalar ratio. In section 4, we study warm inflation with  Natural potential in weak and strong dissipative regimes for variable dissipation coefficient. In section 5, we present our conclusions.

\section{Overview of $f(\phi)T$ gravity}    
 
 The action in $f(\phi)T$ gravity theory as proposed in \cite{r26} is given by,

\begin{equation}
     S = \int \left[\frac{R}{2\kappa} + \lambda f(\phi) T + L_m \right] \sqrt{-g} d^4x
     \label{e18}
\end{equation}
 where a coupling term (coupling between inflaton $(\phi)$ and trace of the energy-momentum tensor $(T)$ is added to the Hilbert-Einstein action. Here, $\kappa=8\pi G=1/ M_P^2$, $M_P$ is the reduced Planck mass, $R$ stands for Ricci scalar, $L_m$ is the matter Lagrangian density, g is the determinant of the metric tensor, $\lambda$ is the model parameter, and $f(\phi)$ is a dimensionless function of the inflaton $\phi$ which satisfies the condition $f(0)=0$. This indicates that when inflaton field decays and its energy no longer dominates the universe, Einstein gravity is recovered. Also in the limit $\lambda\rightarrow 0$, the action in equation \eqref{e18} returns to the action in the Einstein theory.  
 
 In this work, we have considered the form of $f(\phi)$ as $f(\phi)=\sqrt{\kappa } \phi$. Also, we have used the natural units such that $c=\hslash=1$ and used the $(-,+,+,+)$ sign convention for the metric tensor throughout this manuscript. With the form of $f(\phi)=\sqrt{\kappa}\phi$, the action reads
\begin{equation}
    S = \int \left[\frac{R}{2\kappa} +  \lambda \sqrt{\kappa } \phi T + L_m \right] \sqrt{-g} d^4x 
    \label{e19}
\end{equation}
 By varying the action \eqref{e19} with respect to the metric, we get the modified Einstein field equation in the following form,
\begin{equation}
    R_{\alpha\beta} - \frac{1}{2}g_{\alpha\beta}R = \kappa T_{\alpha\beta}^{(eff)}
    \label{e22}
\end{equation}
where $T_{\alpha\beta}^{(eff)}$ is the effective stress-energy tensor given by,
\begin{equation}
    T_{\alpha\beta}^{(eff)}= T_{\alpha\beta} - 2 \lambda \sqrt{\kappa } \phi( T_{\alpha\beta} - \frac{1}{2}T g _{\alpha\beta} + \Theta_{\alpha\beta})
    \label{a}
\end{equation}
where $T$ is the trace of the energy-momentum tensor $T_{\alpha\beta}$ in Einstein gravity. In flat FLRW background $T_{\alpha\beta}$ is given by

\begin{equation}
    T_{\alpha\beta} = \partial_{\alpha}\phi \partial_{\beta}\phi + g_{\alpha\beta} \left(\frac{1}{2}\dot \phi^2 - V(\phi)\right)
    \label{j}
\end{equation}
where $V(\phi)$ is the potential of the inflaton. For inflaton field, $\Theta_{\alpha\beta}$ reads as
\begin{equation}
    \Theta_{\alpha\beta}=-\partial_{\alpha}\phi \partial_{\beta}\phi-T_{\alpha\beta}
    \label{b}
\end{equation}
From equations \eqref{a},\eqref{j} and \eqref{b}, the effective energy density and pressure for inflaton field are obtained as
\begin{equation}
    T_{00}^{(eff)}= \rho_\phi^{(eff)}=(1+2\lambda \sqrt{\kappa}\phi)\frac{\dot{\phi}^2}{2}+(1+4\lambda \sqrt{\kappa}\phi)V
    \label{e20}
 \end{equation}
\begin{equation}
   T_{ij}^{(eff)}=  P_\phi^{(eff)} g_{ij}=\left[(1+2\lambda \sqrt{\kappa}\phi)\frac{\dot{\phi}^2}{2}-(1+4\lambda \sqrt{\kappa}\phi)V\right]g_{ij}
   \label{e21}
 \end{equation}
 
Substituting equation \eqref{e20} into the $00$ component of the modified Einstein field equation \eqref{e22}, we get
\begin{equation}
    H^2 = \frac{\kappa}{3}\left[\frac{\dot\phi^2}{2} (1+ 2\lambda \sqrt{\kappa } \phi) + V (1+ 4\lambda \sqrt{\kappa } \phi)\right]
\end{equation}
which is known as the modified Friedmann equation. Here we have defined the Hubble parameter as $H=\frac{\dot a}{a}$. Similarly, substituting equation \eqref{e21} into the $ij$ component of the modified Einstein field equation \eqref{e22}, we get
\begin{equation}
    \frac{\ddot a}{a} = - \frac{\kappa}{3} \left[\dot\phi^2 (1+ 2\lambda \sqrt{\kappa } \phi) - V (1+ 4\lambda \sqrt{\kappa } \phi)\right]
    \label{e23}
\end{equation}
This is the modified acceleration equation. We can also obtain the expression for $\dot H$ as,
\begin{equation}
    \dot H = \frac{\ddot a}{a} - H^2 = - \frac{\kappa}{2} (p^{eff} + \rho^{eff})= -\frac{\kappa}{2}\left[\dot\phi^2(1+ 2\lambda \sqrt{\kappa } \phi)\right]
    \label{e24}
\end{equation}
 Equations \eqref{e23} and \eqref{e24} are known as the modified Friedmann second equation. In the limit $\lambda=0$, these equations reduce to those in Einstein gravity.
 
Also the continuity equation or the modified Klein-Gordon equation can be written as,
\begin{equation}
    \ddot \phi (1+ 2\lambda \sqrt{\kappa } \phi) + 3H\dot \phi (1+ 2\lambda \sqrt{\kappa } \phi) +\lambda \sqrt{\kappa } \dot{\phi}^2 +   (1+ 4\lambda \sqrt{\kappa } \phi)  V^\prime+4\lambda \sqrt{\kappa } V = 0
\end{equation}
 where the prime denotes derivative with respect to the inflaton field.

 \section{Warm Inflation in  $f(\phi)T$ gravity}
  In the warm inflationary paradigm, the inflaton field interacts with other fields during the inflationary period, leading to the decay of the inflaton field into radiation. To capture this feature, a dissipation co-efficient $\Gamma$ is added to the equation of motion describing the warm inflationary scenario. The equations that completely describe the dynamics of the warm inflation scenario in $ f(\phi)T$ gravity can be described as  
\begin{equation}
    H^2=\frac{\kappa}{3}(\rho_\phi^{(eff)}+\rho_\gamma)
    \label{e25}
\end{equation}
\begin{equation}
\dot{\rho}_\phi^{(eff)}+3H(\rho_\phi^{(eff)}+P_\phi^{(eff)})=-\Gamma \dot{\phi}^2
\label{e29}
\end{equation}
\begin{equation}
    \dot{\rho}_\gamma+4H\rho_\gamma=\Gamma \dot{\phi}^2
    \label{e31}
\end{equation}
Here $\rho_\phi^{(eff)}$, $\rho_\gamma$ and $P_\phi^{(eff)}$ are the energy density of the scalar field, energy density of the radiation field and pressure of the scalar field respectively. $\Gamma $ is the dissipation coefficient which describes the decay of inflaton into radiation during inflationary phase.

Modified equation of motion of the inflaton field in warm inflation can be derived by substituting equations \eqref{e20} and \eqref{e21} into equation \eqref{e29}, 
 \begin{equation}
    \ddot \phi (1+ 2\lambda \sqrt{\kappa } \phi) + 3H\dot \phi (1+ 2\lambda \sqrt{\kappa } \phi+Q) +\lambda \sqrt{\kappa } \dot{\phi}^2 + (1+ 4\lambda \sqrt{\kappa } \phi) V^\prime+4\lambda \sqrt{\kappa } V = 0
    \label{e30}
 \end{equation}
 where $Q=\frac{\Gamma}{3H}$ is the dissipation rate which describes the effectiveness at which the inflaton energy converts into radiation. The $Q\gg1$ corresponds to strong dissipation regime, and the $Q\ll1$ corresponds to weak dissipation regime \cite{r29}. The presence of extra friction term in the equation of motion of the inflaton suggests that radiation will not be redshifted during inflation, because, inflaton field continuously converts into radiation through dissipation

 For inflation to occur and last long enough, the potential energy of the inflaton field must be dominated over both the kinetic energy of the inflaton field and the energy density of the radiation field, however, the radiation energy density still satisfies the condition $\rho_\gamma^{1/4}>H$. That means $T>H$ during warm inflation and this condition is considered as the condition for warm inflation. It is also assumed that the production of radiation is quasi-stable. These approximations are called slow-roll approximations which can be quantified as

 \begin{equation}
     \frac{(1+2\lambda \sqrt{\kappa}\phi)}{2}\dot{\phi}^2+\rho_\gamma<<(1+4\lambda \sqrt{\kappa}\phi)V
\end{equation}
 \begin{equation}
    (1+2\lambda \sqrt{\kappa}\phi)\ddot \phi << 3H\dot \phi (1+ 2\lambda \sqrt{\kappa}\phi+Q) 
 \end{equation}
 \begin{equation}
    \dot{\rho}_\gamma<< 4H\rho_\gamma 
\end{equation}
\begin{equation}
     \sqrt{\kappa}\dot{\phi}^2<<H\dot{\phi}
\end{equation}
 Under these considerations, $\rho_\phi\simeq V(\phi)$, and the dynamical equations \eqref{e25}, \eqref{e30} and \eqref{e31} read as
 \begin{equation}
    H^2=\frac{ \kappa}{3}(1+4\lambda \sqrt{\kappa}\phi)V
    \label{e8}
\end{equation}
 \begin{equation}
       3H\dot \phi (1+ 2\lambda \sqrt{\kappa } \phi+Q) + (1+ 4\lambda \sqrt{\kappa } \phi)V^\prime+4\lambda \sqrt{\kappa } V = 0
       \label{e32}
 \end{equation}
 \begin{equation}
     \rho_\gamma=\frac{\Gamma \dot{\phi}^2}{4H} =C\tau^4
     \label{e9}
 \end{equation}
 where $C=\frac{g_*\pi^2}{30}$ is the Stefan-Boltzmann constant and $g_*$ is the number of degrees of freedom for the radiation at temperature $\tau$. (For calculation we will take $C = 70$ for $ g_*= 200$)

The slow-roll approximation can be parameterized by several slow-roll parameters namely $\epsilon$, $\eta$, and $\beta$. Using equations \eqref{e8}, \eqref{e32} and \eqref{e9}, we can derive the modified slow-roll parameters in the framework of $f(\phi)T$ gravity in terms of the inflaton potential $V(\phi)$ as

 \begin{eqnarray}
\epsilon&=&-\frac{\dot{H}}{H^2}  \nonumber\\
&=&\frac{1}{2\kappa(1+2\lambda \sqrt{\kappa}\phi+Q)} \left(\frac{V^\prime}{V}+\frac{4\lambda \sqrt{\kappa}}{1+4\lambda \sqrt{\kappa}\phi}\right)^2
 \label{79}
 \end{eqnarray}
  
  \begin{eqnarray}
  \eta &=&-\frac{\ddot{\phi}}{H\dot{\phi}}\nonumber \\
  &=&\frac{1}{\kappa(1+2\lambda \sqrt{\kappa}\phi+Q)} \left(\frac{V^{\prime\prime}}{V}+\frac{\lambda \sqrt{\kappa}(6+8\lambda \sqrt{\kappa}\phi+8Q)}{(1+4\lambda \sqrt{\kappa}\phi)(1+2\lambda \sqrt{\kappa}\phi+Q)}\frac{V^\prime}{V}\nonumber\right.\\&-&\left.\frac{8\lambda^2\kappa}{(1+2\lambda \sqrt{\kappa}\phi+Q)(1+4\lambda \sqrt{\kappa}\phi)}\right) 
  \label{80}
  \end{eqnarray}
 and
  \begin{eqnarray}
  \beta&=&-\frac{\dot{\rho}_\gamma}{H\rho_\gamma} \nonumber \\&=& \frac{1}{\kappa(1+2\lambda \sqrt{\kappa}\phi+Q)}\frac{\Gamma^\prime}{\Gamma} \left(\frac{V^\prime}{V}+\frac{4\lambda \sqrt{\kappa}}{1+4\lambda \sqrt{\kappa}\phi}\right)
 \label{81}
  \end{eqnarray}
  The conditions satisfied by the slow-roll parameters for warm inflation to occur are $\epsilon \ll 1$, $|\eta| \ll 1$ and $|\beta| \ll 1$ \cite{r39}. Violation of the slow-roll conditions marks the end of inflation.

 Another significant parameter during inflation is the number of e-folding. This parameter quantifies the extent of expansion of the universe throughout the inflationary epoch. The number of e-foldings when the inflation field $\phi$ rolls from its value $\phi_i$ to $\phi_f$ is estimated as:

 \begin{eqnarray}
N&=&\int^{t_2}_{t_1} H dt=\int^{\phi_f}_{\phi_i}\frac{H}{\dot{\phi}}d\phi\nonumber\\ &=& -\kappa\int^{\phi_f}_{\phi_i}\frac{(1+2\lambda \sqrt{\kappa}\phi+Q)}{\frac{V^\prime}{V}+\frac{4\lambda \sqrt{\kappa}}{1+4\lambda \sqrt{\kappa}\phi}}d\phi
\label{115}
\end{eqnarray}

\subsection{Perturbation Spectra}
In this section, we develop the theory of cosmological perturbations for warm inflation in the $f(\phi)T$ theory of gravity. In warm inflationary models $T>H$, the fluctuations of the inflaton field will be produced by thermal effects due to radiation. Consequently, the source of density fluctuations in warm inflationary models is the thermal fluctuations in the radiation field rather than quantum fluctuations. These thermal fluctuations in the radiation field produce fluctuations in the inflaton field. These fluctuations lead to perturbations in the metric which in turn perturb the equation of motion of inflaton field. In the following, we will calculate the fluctuations of the inflaton field. To compute the inflaton fluctuation we start with the perturbations of the FRW metric in the spatially flat gauge
which is given by
\begin{equation}
      ds^2=-(1+2A)dt^2+2a(t)\partial_iBdtdx^i+a^2(t)\delta_{ij}dx^idx^j
      \label{200}
\end{equation}
 Where $A(x,t)$ and $B(x,t)$ are spece-time dependent scalar perturbations of the metric. Under this perturbation, we expand the inflaton field $\phi(x,t)= \phi(t)+\delta\phi(x,t)$, where $\delta\phi(x,t)$ is the linear response due to the thermal stochastic noise. Using the slow-roll conditions, we find the perturbed equation of the inflaton field in the momentum space given by,
 \begin{eqnarray}
 &&(1+2 \lambda \sqrt{\kappa}\phi)\ddot{\delta\phi}_k+3H(1+2\lambda \sqrt{\kappa}\phi+Q)\dot{\delta\phi}_k+\frac{k^2}{a^2}(1+2\lambda \sqrt{\kappa}\phi)\delta\phi_k=(1+2\lambda \sqrt{\kappa}\phi)\dot{\phi}\dot{A}\nonumber
 \\&+&\frac{k^2}{a}(1+2\lambda \sqrt{\kappa}\phi)\dot{\phi}B-(2(1+4\lambda \sqrt{\kappa}\phi)V^\prime+8\lambda \sqrt{\kappa} V+\Gamma\dot{\phi})A
 \label{201}
 \end{eqnarray}
 And the perturbed Einstein equation becomes
 \begin{equation}
      3H^2A+\frac{k^2}{a}HB=-4\pi G\delta\rho
      \label{e33}
  \end{equation}
  \begin{equation}
      HA=4\pi G (\rho+p)\delta u 
      \label{203}
  \end{equation}
  In the above equations, $\rho$ and $p$ denote the total energy density and pressure respectively, $\delta\rho$ denotes the perturbations at the linear order of the total energy density, and $\delta u$ is the scalar part of the linear perturbation of the 4- velocity. Substituting the expressions of $A$ and $B$ from equations \eqref{e33} and \eqref{203} into equation \eqref{201}, we get the perturbed equation of motion of inflation field as
\begin{eqnarray}
 &&(1+2\lambda \sqrt{\kappa}\phi)\ddot{\delta\phi}_k(t)+3H(1+2\lambda \sqrt{\kappa}\phi+Q)\dot{\delta\phi}_k(t)+\frac{k^2}{a^2}(1+2\lambda \sqrt{\kappa}\phi)\delta\phi_k(t)=\xi_k(t) 
 \label{205}
 \end{eqnarray}
where the thermal stochastic noise source $\xi_k(t)$ is introduced to describe thermal fluctuations. In the slow-roll regime, the term $ \ddot{\delta\phi}_k$ in equation \eqref{205} can be neglected. Thus equation \eqref{205} becomes
\begin{eqnarray}
 && 3H(1+2\lambda \sqrt{\kappa}\phi+Q)\dot{\delta\phi}_k(t)+\frac{k^2}{a^2}(1+2\lambda \sqrt{\kappa}\phi)\delta\phi_k(t)=\xi_k(t) 
 \label{206}
 \end{eqnarray}
The solution of equation \eqref{206} is
  
 \begin{equation}
   \delta\phi_k(t)=\frac{1}{3H(1+2\lambda \sqrt{\kappa}\phi+Q)}\exp{\left[-\frac{t}{\mathcal{T}}\right]} \int_{t_0}^t\exp{\left[\frac{t^\prime}{\mathcal{T}}\right]}\xi_k(t^\prime)dt^\prime+ \delta\phi_k(t_0)\exp{\left[-\frac{t-t_0}{\mathcal{T}}\right]}
   \label{e34}
 \end{equation}
 where,
  $ \mathcal{T}(\phi)=\frac{3H (1+2\lambda \sqrt{\kappa}\phi+Q)}{(1+2\lambda \sqrt{\kappa}\phi) \frac{k^2}{a^2}}= \frac{3H (1+2\lambda \sqrt{\kappa}\phi+Q)}{(1+2\lambda \sqrt{\kappa}\phi)k_p^2}$ and $k_p$ is the physical wave number. The first term in the right-hand side of the solution \eqref{e34} is the noise contribution which acts to thermalize $\delta\phi$. whereas the second term is the memory term for the initial value of $\delta\phi$ which is exponentially damping i.e. it becomes negligible with time.  If $k_p$ of one mode of $\delta\phi_k$  is smaller than the freeze-out physical wave number $k_F$, then the mode will not thermalize during the Hubble time, and for $k_p\gtrsim k_F$, the mode will be thermalized and the memory term will vanish within Hubble time. So, the freeze-out wave number $k_F$ is   
 \begin{equation}
     k_F=\sqrt{\frac{3H^2(1+2\lambda \sqrt{\kappa}\phi+Q)}{1+2\lambda \sqrt{\kappa}\phi} }
     \label{207}
 \end{equation}
The power spectrum for the scalar fluctuations in warm inflation has the form 
 \begin{equation}
     P_R=\left(\frac{H}{\dot{\phi}}\right)^2\delta\phi^2
     \label{116}
 \end{equation}
where the fluctuations of scalar field can be obtained through the relation

\begin{eqnarray}
 \delta\phi^2&=&\frac{k_F \tau}{2\pi^2}
 \label{20}
 \end{eqnarray}
Combining equations \eqref{207},\eqref{116} and \eqref{20}, the expression for scalar power spectrum $P_R$ for warm inflation in $f(\phi)T$ gravity leads to
\begin{eqnarray}
P_R &=&\frac{H^3 \tau}{2\pi^2 \dot{\phi}^2}\sqrt{\frac{3(1+2\lambda \sqrt{\kappa}\phi+Q)}{1+2\lambda \sqrt{\kappa}\phi}}
\label{117}
\end{eqnarray}
In the limit $\lambda\rightarrow 0$, the above expression goes back to the form of $P_R$ for warm inflation in Einstein gravity.

The power spectrum for the tensor perturbation is
\begin{eqnarray}
P_T&=&\frac{16 H^2}{\pi M_P^2}\nonumber \\
&=& \frac{16 (1+4\lambda \sqrt{\kappa}\phi)V}{3\pi M_P^4}
\label{118}
\end{eqnarray}
The spectral index $n_s$ and tensor to scalar ratio $r$ are defined as \cite{r41, r42}
\begin{equation}
  n_s-1=\frac{d \ln P_R}{d \ln k}
  \label{e16}
\end{equation}
\begin{equation}
    r=\frac{ P_T}{P_R}
    \label{e17}
\end{equation}
where $P_R$  is the scalar power spectrum and $P_T $ is the tensor power spectrum \cite{r39}.

\section{Warm Natural Inflation in $f(\phi)T$ Gravity with Temperature-Dependent Dissipation Coefficient}

  A successful inflationary model must generate sufficient expansion to solve the horizon problem. To satisfy this constraint, the slope of the inflaton potential must be very flat. This imposes a restriction on the choice of potentials. Moreover, it is expected that such a potential should be motivated by a fundamental theory. The natural inflation model is a well-motivated model where axion plays the role of inflaton field and the shift symmetry present in the axionic theory assures the flat potential required for inflation. We, therefore, attempt to study 
 natural warm inflation in the framework of $f(\phi, T)$ gravity. The form of the natural potential we are considering is given by \cite{r43, r44, r45} 
\begin{equation}
  V(\phi)= \mu^4\left(1+\cos\left(\frac{\phi}{\mathfrak{f}}\right)\right)  
  \label{e10}
\end{equation}
where $\mathfrak{f}$ is the decay constant and $\mu$ is the mass scale of the axion. We assume that the inflaton couples with light gauge fields and produces thermal friction which is given by the dissipation coefficient of the  form \cite{r46, r47, r48}
\begin{equation}
    \Gamma(\tau)=C_\Gamma\frac{\tau^3}{\mathfrak{f}^{2}}
    \label{e11}
\end{equation}
where $C_\Gamma$ is a dimensionless factor proportional to the coupling constant between the inflaton field $\phi$ and gauge field.
  
With this form of dissipation coefficient, we carry out our further analysis in two dissipative regimes viz weak $Q<<1$ and strong $Q>>1$.

  \subsection{Weak Dissipative Regime}
   
  In weak dissipative regime, $Q<<1$ and  the slow-roll parameters $\epsilon$, $\eta$ and $\beta$ take the forms,
 
\begin{equation}
    \epsilon=\frac{1}{2 \kappa (1+2\lambda \sqrt{\kappa}\phi)}\left(\frac{4 \lambda \sqrt{\kappa} }{4 \lambda \sqrt{\kappa} \phi +1}-\frac{\sin \left(\frac{\phi }{\mathfrak{f}}\right)}{\mathfrak{f} \left(\cos \left(\frac{\phi }{\mathfrak{f}}\right)+1\right)}\right)^2
\end{equation}
\begin{eqnarray}
    \eta &=&\frac{1}{\kappa (1+2\lambda \sqrt{\kappa}\phi)} \left(-\frac{8 \lambda ^2 \kappa}{(2 \lambda \sqrt{\kappa} \phi +1) (4 \lambda \sqrt{\kappa} \phi +1)}-\frac{\cos \left(\frac{\phi }{\mathfrak{f}}\right)}{\mathfrak{f}^2 \left(\cos \left(\frac{\phi }{\mathfrak{f}}\right)+1\right)}\nonumber \right.\\  
    &-&\left.\frac{\lambda \sqrt{\kappa} (8 \lambda \sqrt{\kappa} \phi +6) \sin \left(\frac{\phi }{\mathfrak{f}}\right)}{\mathfrak{f} (2 \lambda \sqrt{\kappa} \phi +1) (4 \lambda \sqrt{\kappa} \phi +1)\left(\cos \left(\frac{\phi }{\mathfrak{f}}\right)+1\right) } \right)
\end{eqnarray}
 \begin{equation}
      \beta= \frac{1}{\kappa(1+2\lambda \sqrt{\kappa}\phi)}\frac{\Gamma^\prime}{\Gamma} \left( \frac{4\lambda \sqrt{\kappa}}{1+4\lambda \sqrt{\kappa}\phi}-\frac{\sin \left(\frac{\phi }{\mathfrak{f}}\right)}{\mathfrak{f} \left(\cos \left(\frac{\phi }{\mathfrak{f}}\right)+1\right)}\right)
  \end{equation}
  
 and the equation of motion of the inflaton field reads
\begin{equation}
       3H\dot \phi (1+ 2\lambda \sqrt{\kappa } \phi) + (1+ 4\lambda \sqrt{\kappa } \phi)V^\prime+4\lambda \sqrt{\kappa } V = 0
 \end{equation} 
 Using equations \eqref{e8} and \eqref{e10}, we obtain
\begin{equation}
    \dot{\phi}=-\frac{\sqrt{\mu ^4 (4 \lambda \sqrt{\kappa} \phi +1) \left(\cos \left(\frac{\phi }{\mathfrak{f}}\right)+1\right)} \left(\frac{4 \lambda \sqrt{\kappa}}{4 \lambda \sqrt{\kappa} \phi +1}-\frac{\sin \left(\frac{\phi }{\mathfrak{f}}\right)}{\mathfrak{f} \left(\cos \left(\frac{\phi }{\mathfrak{f}}\right)+1\right)}\right)}{\sqrt{3 \kappa} (2 \lambda \sqrt{\kappa} \phi +1)}
    \label{e14}
\end{equation}
 The temperature of the thermal bath can be obtained by using equation \eqref{e9} as
 \begin{equation}
     \tau=\left(  \frac{3 Q \dot{\phi}^2}{4C}\right)^\frac{1}{4}
     \label{202}
 \end{equation}
 Also, the relation \eqref{e11}, reads
 \begin{equation}
     \tau=\left(\frac{3Q H \mathfrak{f}^2}{C_\Gamma}\right)^\frac{1}{3}
     \label{e12}
 \end{equation}
Equating equations \eqref{202} and \eqref{e12}, we get
\begin{equation}
    Q= \frac{\left(\frac{1}{4 C}\right)^3 \left(C_\Gamma^4 \mu ^4 \right) \left((4 \lambda \sqrt{\kappa} \phi +1) \left(\cos \left(\frac{\phi }{\mathfrak{f}}\right)+1\right)\right) \left(\frac{4 \lambda \sqrt{\kappa} \mathfrak{f}}{4 \lambda \sqrt{\kappa} \phi +1}-\frac{\sin \left(\frac{\phi }{\mathfrak{f}}\right)}{\cos \left(\frac{\phi }{\mathfrak{f}}\right)+1}\right)^6}{\left(9 \kappa^5 \mathfrak{f}^{14}\right) (2 \lambda \sqrt{\kappa} \phi +1)^6}  
    \label{e15}
\end{equation}

The violation of the slow-roll conditions governs the end of inflation. It is numerically checked that $\epsilon$ violates the slow-roll condition first. So, by using the condition $\epsilon(\phi_{f})=1$, we determine the value of the scalar field at the end of inflation. The number of e-foldings (equation \eqref{115}) in this regime becomes,  

\begin{eqnarray}
N&=& - \kappa\int^{\phi_f}_{\phi_i}\frac{(1+2\lambda \sqrt{\kappa}\phi)}{\frac{V^\prime}{V}+\frac{4\lambda \sqrt{\kappa}}{1+4\lambda \sqrt{\kappa}\phi}}d\phi
\label{e111}
\end{eqnarray}

Using equation \eqref{e111}, the initial field value $(\phi_i)$ is calculated for $N=60$. The temperature of the thermal bath can be obtained by inserting the equations \eqref{e14} and \eqref{e15} into equations \eqref{202},

\begin{equation}
   \tau= 0.144 \sqrt[4]{\frac{C_\Gamma^4 \mu ^8 (4 \lambda \sqrt{\kappa}  \phi +1)^2 \left(\cos \left(\frac{\phi }{\mathfrak{f}}\right)+1\right)^2      \left(\frac{4 \lambda \sqrt{\kappa} \mathfrak{f}}{4 \lambda \sqrt{\kappa}  \phi +1}-\frac{\sin \left(\frac{\phi }{\mathfrak{f}}\right)}{\cos \left(\frac{\phi }{\mathfrak{f}}\right)+1}\right)^6     \left(\frac{4 \lambda \sqrt{\kappa} }{4 \lambda \sqrt{\kappa}  \phi +1}-\frac{\sin \left(\frac{\phi }{\mathfrak{f}}\right)}{\mathfrak{f} \left(\cos \left(\frac{\phi }{\mathfrak{f}}\right)+1\right)}\right)^2}{\kappa^5 C^4 \mathfrak{f}^{14}(2 \lambda \sqrt{\kappa}  \phi +1)^8}} 
\end{equation}
  From the above relation, we can obtain the temperature of the thermal bath for any value of $\phi$. Also, by substituting the relations \eqref{117}, \eqref{118}, \eqref{e14} and \eqref{e15} into equations \eqref{e16} and \eqref{e17}, we obtain the spectral index and tensor-to-scalar ratio at initial field value $\phi_i$.

  To evaluate the viability of any inflationary model, it is necessary to check whether the values of cosmological observables are in good agreement with the observational data. So, in order to constrain different parameters used in our model, we consider the latest limits set by Planck 2018 and BICEP/Keck data on $n_s$ and $r$.

\begin{figure}[h!]
\centering
 \begin{subfigure}[h]{0.32\textwidth}
     \centering
     \includegraphics[width=\textwidth]{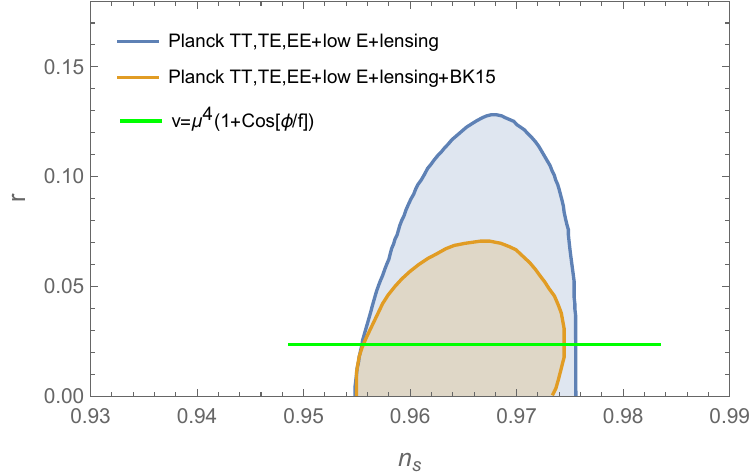}
     \caption{$C_\Gamma=1.1\times 10^{5}$}
 \end{subfigure}
 \hfill
 \begin{subfigure}[h]{0.32\textwidth}
    \centering
    \includegraphics[width=\textwidth]{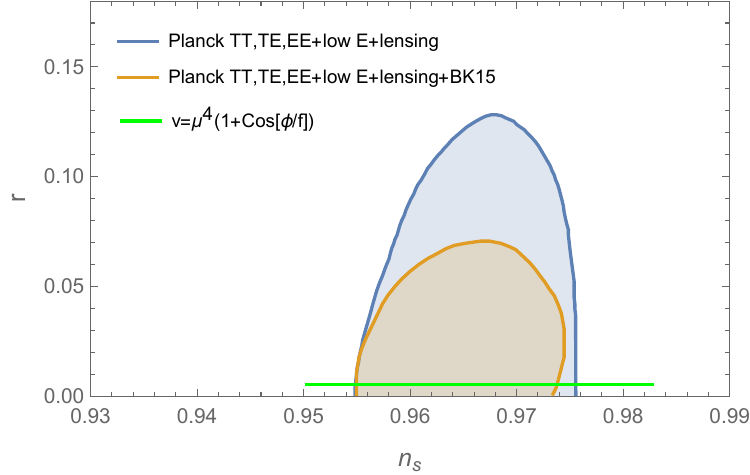}
     \caption{$C_\Gamma=5\times 10^{5}$}
\end{subfigure}
\hfill
 \begin{subfigure}[h]{0.32\textwidth}
     \centering
     \includegraphics[width=\textwidth]{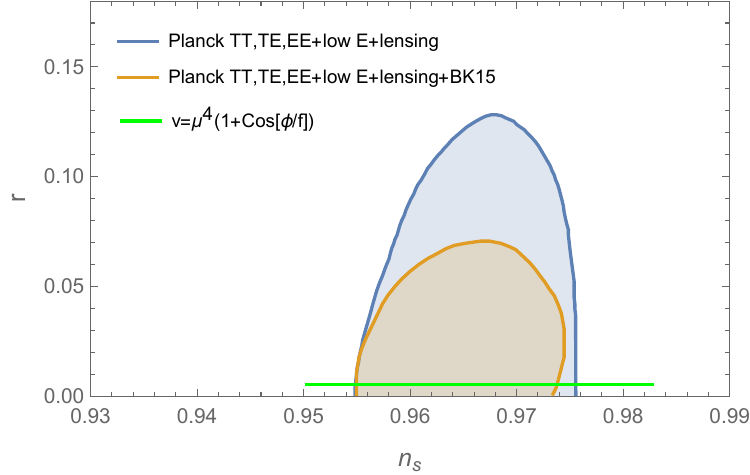}
     \caption{$C_\Gamma=7.38\times 10^{5}$}
      
\end{subfigure}
 
    \caption{  The spectral index $n_s$, and the tensor-to-scalar ratio $r$ predicted by warm natural inflationary model in the $f(\phi)T$ gravity theory for $N=60$ in weak dissipative regime (green solid line). The marginalized joint $95\%$ CL regions for the spectral index $n_s$, and the tensor-to-scalar ratio $r$ from Planck 2018 data alone, and in combinations with BICEP/Keck Array are shown in blue and orange respectively.}
    \label{f4}
\end{figure}

  The predictions of the spectral tilt and tensor-to-scalar ratio for three representative values of dissipation parameters are shown in  Fig.~\ref{f4}. The plots are obtained by varying $\lambda$ and fixing the values of the mass scale $\mu=0.001 M_P$ and axion decay constant $\mathfrak{f}=0.4M_P$ for 60 e-foldings. It is seen that variation of tensor-to-scalar ratio is very slow and with the increasing dissipation parameter $(C_\Gamma)$, its values go on decreasing. The ranges of model parameter $\lambda$ are found to be $[93.6, 97.3]$ for $C_\Gamma=1.1\times10^5$, $[93.57, 97.27]$ for $C_\Gamma=5\times10^5$ and $[93.47, 97.17]$ for $C_\Gamma=7.38\times10^5$. The plot shows that the warm natural inflationary scenario in the $f(\phi)T$ gravity framework is consistent with observational constraints on $r$ and $n_s$ in weak dissipative regime. 

\subsection{Strong Dissipative Regime}
   
  In strong dissipative regime, $Q>>1$. So, the slow-roll parameters $\epsilon$, $\eta$ and $\beta$ in this regime become,
 
\begin{equation}
    \epsilon=\frac{1}{2 \kappa (2\lambda \sqrt{\kappa}\phi +Q)}\left(\frac{4 \lambda \sqrt{\kappa} }{4 \lambda \sqrt{\kappa} \phi +1}-\frac{\sin \left(\frac{\phi }{\mathfrak{f}}\right)}{\mathfrak{f} \left(\cos \left(\frac{\phi }{\mathfrak{f}}\right)+1\right)}\right)^2
    \label{130}
\end{equation}
\begin{eqnarray}
    \eta &=&\frac{1}{ \kappa (2\lambda \sqrt{\kappa}\phi+Q)} \left(-\frac{8 \lambda ^2 \kappa}{(2 \lambda \sqrt{\kappa} \phi +Q) (4 \lambda \sqrt{\kappa} \phi +1)}-\frac{\cos \left(\frac{\phi }{\mathfrak{f}}\right)}{\mathfrak{f}^2 \left(\cos \left(\frac{\phi }{\mathfrak{f}}\right)+1\right)}\nonumber \right.\\  
    &-&\left.\frac{\lambda \sqrt{\kappa} (8 \lambda \sqrt{\kappa} \phi +6+8Q) \sin \left(\frac{\phi }{\mathfrak{f}}\right)}{\mathfrak{f} (2 \lambda \sqrt{\kappa} \phi +Q) (4 \lambda \sqrt{\kappa} \phi +1)\left(\cos \left(\frac{\phi }{\mathfrak{f}}\right)+1\right) } \right)
    \label{131}
\end{eqnarray}
 \begin{equation}
      \beta= \frac{1}{\kappa(2\lambda \sqrt{\kappa}\phi+Q)}\frac{\Gamma^\prime}{\Gamma} \left( \frac{4\lambda \sqrt{\kappa}}{1+4\lambda \sqrt{\kappa}\phi}-\frac{\sin \left(\frac{\phi }{\mathfrak{f}}\right)}{\mathfrak{f} \left(\cos \left(\frac{\phi }{\mathfrak{f}}\right)+1\right)}\right)
      \label{132}
  \end{equation}
 and the equation of motion of the inflaton field becomes
\begin{equation}
       3H\dot \phi ( 2\lambda \sqrt{\kappa } \phi+Q) + (1+ 4\lambda \sqrt{\kappa } \phi) V^\prime+4\lambda \sqrt{\kappa } V = 0
       \label{133}
 \end{equation} 
  From $\Gamma(\tau)=C_\Gamma \frac{\tau^3}{\mathfrak{f}^2} $, we have $\tau=\left(\frac{3Q \mathfrak{f}^2 H}{C_\Gamma}\right)^\frac{1}{3} $. Equating this with $ \tau=\left(  \frac{3 Q \dot{\phi}^2}{4C}\right)^\frac{1}{4}$, we get
 \begin{equation}
     \dot{\phi}^6=3 Q (4 C)^3 \left(\frac{\mathfrak{f}^2 H}{C_\Gamma}\right)^4
     \label{134}
 \end{equation}
 Expressions of $\dot{\phi}$ and $Q$ can be obtained by solving equations \eqref{133} and \eqref{134}. These expressions are substituted in equations \eqref{130}, \eqref{131} and \eqref{132} to obtain the slow-roll parameters $\epsilon$, $\eta$ and $\beta$ in terms of model parameters. In this case, we have numerically checked that inflation ends when $\epsilon$ reaches unity and found the corresponding value of inflaton field $(\phi_f)$. The number of e-foldings (equation \eqref{115}) in this regime becomes,  

\begin{eqnarray}
N&=& -\kappa\int^{\phi_f}_{\phi_i}\frac{(2\lambda \sqrt{\kappa}\phi+Q)}{\frac{V^\prime}{V}+\frac{4\lambda \sqrt{\kappa}}{1+4\lambda \sqrt{\kappa}\phi}}d\phi
\label{e13}
\end{eqnarray}
 
 Using the above definition of $N$, we obtain the initial field value $(\phi_i)$ for 60 e-foldings. From the expressions of the power spectrum $P_R$ (equation \eqref{117}) and $P_T$ (equation \eqref{118}), the spectral index $n_s$ and tensor-to-scalar ratio $r$ are computed at $\phi_i$ for different values of model parameters.
 
\begin{figure}[h!]
\centering
 \begin{subfigure}[h]{0.32\textwidth}
     \centering
     \includegraphics[width=\textwidth]{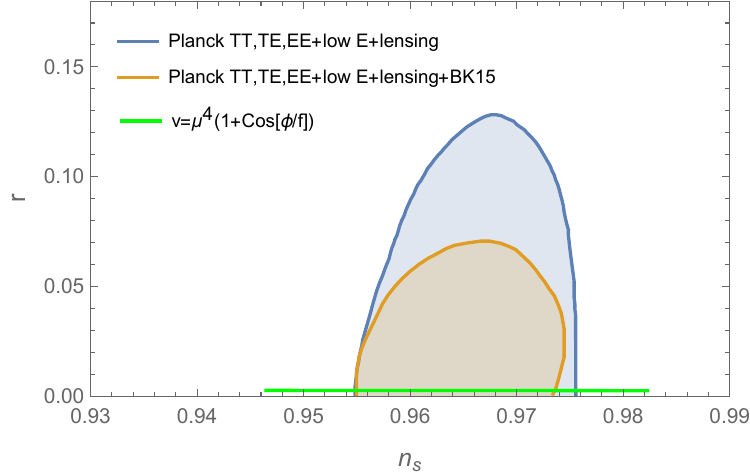}
     \caption{$C_\Gamma= 10^{6}$}
 \end{subfigure}
 \hfill
 \begin{subfigure}[h]{0.32\textwidth}
    \centering
    \includegraphics[width=\textwidth]{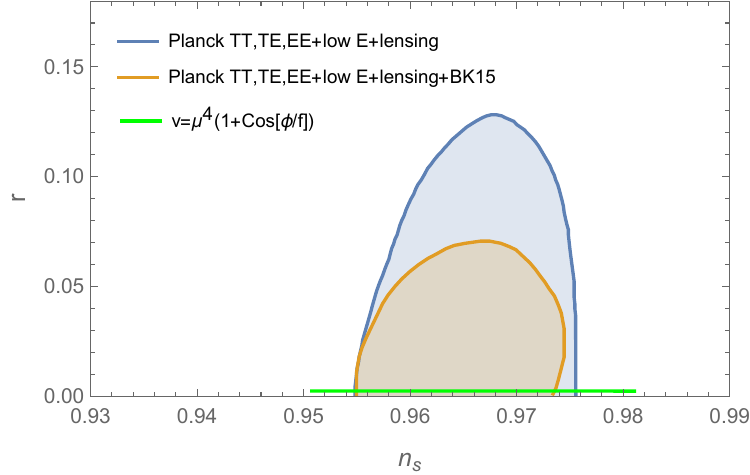}
     \caption{$C_\Gamma=1.1\times 10^{6}$}
\end{subfigure}
\hfill
 \begin{subfigure}[h]{0.32\textwidth}
     \centering
     \includegraphics[width=\textwidth]{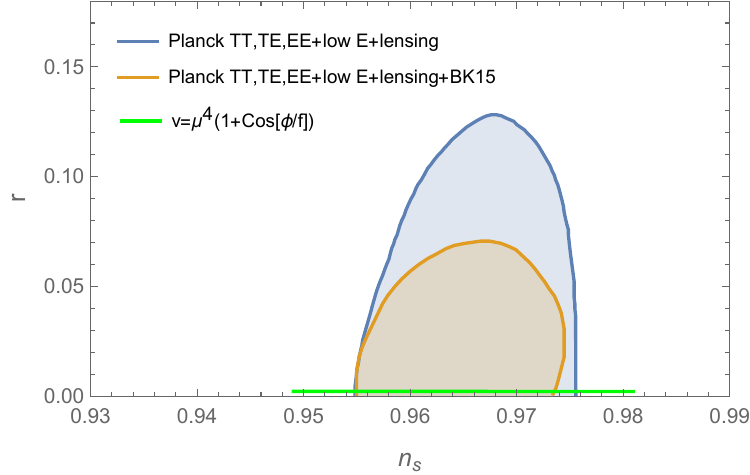}
     \caption{$C_\Gamma=1.2\times 10^{6}$}
      
\end{subfigure}
 
    \caption{  The spectral index $n_s$, and the tensor-to-scalar ratio $r$ predicted by warm natural inflationary model in the $f(\phi)T$ gravity theory for $N=60$ in strong dissipative regime (green solid line). The marginalized joint $95\%$ CL regions for the spectral index $n_s$, and the tensor-to-scalar ratio $r$ from Planck 2018 data alone, and in combinations with BICEP/Keck Array are shown in blue and orange respectively.}
    \label{f6}
\end{figure}

The depictions of the spectral tilt and tensor-to-scalar ratio are shown in  Fig.~\ref{f6}. Here also we consider three representative values of $C_\Gamma$. In each case, the mass scale and the axion decay constant are set at $\mu=0.001 M_P$, $\mathfrak{f}=0.4M_P$. The range of model parameter $\lambda$ is found to be $[88.3, 95.1]$ for $C_\Gamma=10^6$, $[95.75, 104.3]$ for $C_\Gamma=1.1 \times 10^6$ and $[103.2, 114]$ for $C_\Gamma=1.2 \times 10^6$. The values of $n_s$ and $r$ are in good agreement with Planck 2018 \cite{r38} and BICEP \cite{r37} data at $95\%$ CL.

\section{Conclusion}
Over the past few decades, extensive investigations have been conducted to understand and model the dynamics of the universe. The standard model of cosmology successfully describes the formation and evolution of the universe. However, this model has some problems like horizon and flatness problems which remain as open issues. In order to overcome these shortcomings, the idea of inflationary cosmology appears necessary in the early era of the universe. Even though general relativity provides remarkably accurate predictions for describing cosmological phenomena, it can not explain the dark sector of the universe. For this reason, investigating alternative theories of gravity arises as a promising path. 

In this work, we explore natural warm inflation in the framework of $f(\phi)T$ gravity theory. In warm inflation, the interactions between scalar and other fields during inflation are taken into account which provides the dissipation term. Here we consider a cubic temperature-dependent dissipation coefficient.
We have investigated this model in two separate dissipative regimes viz weak and strong and in each case, we have evaluated the cosmological observables for certain ranges of model parameters.

In the weak dissipative regime, the cosmological observables are functions of model parameter $\lambda$, axion decay constant $\mathfrak{f}$, mass scale $\mu$, and dissipation coefficient. We have presented the predictions of the spectral tilt and tensor-to-scalar ratio for dissipation parameter $C_\Gamma=1.1\times10^5$, $C_\Gamma=5\times10^5$ and $C_\Gamma=7.38\times10^5$ by varying $\lambda$ and keeping the other parameters fixed.  The obtained value for the spectral index and the tensor-to-scalar ratio in the selected range of the model parameter $\lambda$  falls in the $1\sigma$ confidence level of the Planck 2018 data and the joint Planck and BICEP results. Hence, we claim that this model in weak dissipative regime is capable of making predictions consistent with the observational data.

In the strong dissipative regime also, the cosmological observables are the function of model parameter $\lambda$, axion decay constant $\mathfrak{f}$, mass scale $\mu$, and dissipation coefficient as expected. Here, we have plotted the predictions of the spectral tilt and tensor-to-scalar ratio for dissipation parameter $C_\Gamma=10^6$, $C_\Gamma=1.1\times10^6$ and $C_\Gamma=1.2\times10^6$ by varying $\lambda$ and keeping the other parameters fixed. The result shows that for some selected ranges of the model parameter $\lambda$, values of the spectral index, and the tensor-to-scalar ratio match in the $1\sigma$ confidence level of the Planck 2018 data and their combination with BICEP/Keck data. Hence, this model is consistent with the observational data in strong disspative regime also.

It is found that the model parameter $\lambda$ has a minimal impact on the tensor-to-scalar ratio, as its values exhibit slow variation. So, it can be concluded that the tensor-to-scalar ratio is less sensitive to the change in the model parameter. The model produces smaller values of tensor-to-scalar ratio in strong dissipative regime. Because, the scalar power spectrum is completely dominated by the dissipation, which strongly suppresses the tensor-to-scalar ratio. Additionally, the value of the axion decay constant can be lowered below the Planck scale which is $0.4 M_P$ in this model. We can realize the natural warm inflation scenario in $f(\phi)T$ gravity for much smaller values of $\mathfrak{f}$ (sub-Planckian scale) compared to the corresponding cold inflation case, which makes it theoretically more sound.

 \section*{Acknowledgement}
 We are thankful to Prof. Sudhakar Panda for fruitful insights.

\end{document}